\documentclass[prl,twocolumn,superscriptaddress]{revtex4-1}
\usepackage[dvips]{graphicx}
\usepackage{type1cm}
\usepackage[usenames,svgnames]{xcolor}
\usepackage{color}
\usepackage{url}
\usepackage{natbib}
\usepackage[colorlinks=true,linkcolor=blue,citecolor=blue,breaklinks=true]{hyperref}

\usepackage{amsmath,amssymb,bm,amsthm, mathtools}
\DeclarePairedDelimiter{\abs}{\lvert}{\rvert}

\DeclareMathOperator{\Tr}{\mathrm{Tr}}
\DeclareMathOperator{\re}{\mathrm{Re}}
\newcommand{\dd}{d}

\newcommand{\ddv}[2]{\frac{d#1}{d#2}}

\usepackage{braket}

\newcommand{\calL}{\mathcal{L}}

\usepackage[capitalize]{cleveref}

\newcommand{\sectionprl}[1]{{\par\it #1.---}}

\begin{document}
\nocite{apsrev41control}

\title{Symmetrized Liouvillian Gap in Markovian Open Quantum Systems}

\author{Takashi Mori}
\affiliation{RIKEN Center for Emergent Matter Science (CEMS), Wako 351-0198, Japan}

\author{Tatsuhiko Shirai}
\affiliation{Department of Computer Science and Communications Engineering, Waseda University, Tokyo 169-8555, Japan}

\begin{abstract}
Markovian open quantum systems display complicated relaxation dynamics.
The spectral gap of the Liouvillian characterizes the asymptotic decay rate towards the steady state, but it does not necessarily give a correct estimate of the relaxation time because the crossover time to the asymptotic regime may be too long.
We here give a rigorous upper bound on the transient decay of auto-correlation functions in the steady state by introducing the symmetrized Liouvillian gap.
The standard Liouvillian gap and the symmetrized one are identical in an equilibrium situation but differ from each other in the absence of the detailed balance condition. 
It is numerically shown that the symmetrized Liouvillian gap always give a correct upper bound on the decay of the auto-correlation function, but the standard Liouvillian gap does not.
\end{abstract}
\maketitle

\sectionprl{Introduction}
It is a fundamental problem in nonequilibrium physics to elucidate how fast a quantum system approaches stationarity under dissipative couplings to an external environment~\citep{Znidaric2015, Sciolla2015, Macieszczak2016, Dann2019, Vernier2020, Mori2020_resolving, Haga2021, Carollo2021}.
This problem is also of great practical interest in quantum technologies.
Because quantum control and computations unavoidably suffer from dissipation and decoherence, it is becoming important to understand general properties of dissipative quantum dynamics~\citep{Lidar2014, Noh2017, Franca2021, Wang2021, Bharti2022}.
Moreover, the strategy of utilizing engineered dissipation in controlling and manipulating quantum states, which was theoretically proposed as reservoir engineering~\citep{Diehl2008, Verstraete2009}, is being implemented in experiments~\citep{Barreiro2011, Barontini2013, Tomita2017}.
It was demonstrated that quantum phase transitions can be induced by controlling the strength of dissipation~\citep{Tomita2017}.
Those ongoing experimental developments will require a more precise theoretical understanding of open-system dynamics.

In the Markovian regime, where the environmental correlation time is much shorter than a typical time of dissipative processes, the dynamics of an open quantum system is generated by the Liouvillian superoperator of the celebrated Lindblad form~\citep{Lindblad1976, Gorini1976}.
One might then expect that the knowledge of the eigenvalue spectrum of the Liouvillian is enough to estimate how fast the relaxation proceeds.
Especially, the Liouvillian gap, which is defined as the smallest nonzero real part of the Liouvillian eigenvalue, has been investigated for various models~\citep{Temme2013, Cai2013, Znidaric2015, Macieszczak2016, Casteels2017, Shibata2019a, Nakagawa2021, Yuan2021, Yoshida2022}.
Because the Liouvillian gap gives the decay rate of the slowest relaxation mode, it is naturally expected that its inverse bounds from above the relaxation time.
However, it turns out that the problem is more elaborate.
It has been shown that the relaxation time is not bounded by the Liouvillian gap~\citep{Mori2020_resolving, Haga2021, Bensa2021, Lee2022}, although the latter characterizes the asymptotic decay rate appearing in the long-time limit~\citep{Kessler2012}.
The point is that the crossover time to the asymptotic regime may be too long especially in many-body systems (it may even diverge in the thermodynamic limit), and hence we should investigate the decay rate \emph{in a transient regime}, which is the main focus of this work.

In this Letter, we provide a rigorous analysis on the decay of auto-correlation functions in the steady state.
Our result tells us that not the standard Liouvillian gap but the symmetrized one appears as a key quantity to bound the relaxation time in the transient regime.
It turns out that the standard Liouvillian gap correctly bounds the relaxation time in an equilibrium situation, but not in a nonequilibrium situation without the detailed balance condition.
In the latter, the symmetrized Liouvillian gap rigorously bounds the relaxation time.

Our results complement a series of results based on the quantum speed limit (QSL), which was originally formulated in isolated quantum systems~\citep{Mandelstam1945} and later extended to open quantum systems~\citep{Deffner2013, DelCampo2013, Marvian2015, Funo2019}.
The QSL gives a \emph{lower} bound on the relaxation time, whereas an \emph{upper} bound is investigated here.

In the following, we first explain the general setup, and then present main results.
We demonstrate the validity of our theoretical results by numerical calculations in an interacting quantum dot coupled to reservoirs.
An extension to time-periodic (i.e. Floquet) systems is briefly mentioned.

\sectionprl{Setup}
Let us consider a Markovian open quantum system, whose state at time $t$ is represented by the density matrix $\rho(t)$.
Its time evolution is generated by the Liouvillian superoperator $\calL$ of the Lindblad form~\citep{Breuer_text}: $\dd\rho(t)/\dd t=-\calL\rho(t)$, where
\begin{align}
    -\calL\rho=-i[\hat{H},\rho]+\sum_k\left(\hat{L}_k\rho\hat{L}_k^\dagger-\frac{1}{2}\left\{\hat{L}_k^\dagger\hat{L}_k,\rho\right\}\right).
    \label{eq:Liouvillian}
\end{align}
The first term of the right-hand side expresses the intrinsic unitary evolution of the system with Hamiltonian $\hat{H}$, whereas the second term represents the dissipation characterized by a set of Lindblad jump operators $\{\hat{L}_k\}$.
The Lindblad form ensures physically natural properties such as the complete positivity~\citep{Lindblad1976, Gorini1976}.

Let us denote by $\{\lambda_\alpha\}$ the eigenvalues of $\calL$.
It is shown that any eigenvalue has a non-negative real part, and hence we sort the eigenvalues in the ascending order:
\begin{align}
    0=\lambda_0<\re\lambda_1\leq\re\lambda_2\leq\dots.
\end{align}
In this work, we assume that the zero eigenvalue is not degenerate: the steady state is unique.
Because of the property $\calL(\rho^\dagger)=\calL(\rho)^\dagger$, it is shown that $\lambda_\alpha^*$ is also an eigenvalue. 
The Liouvillian gap $g$ is defined as
\begin{align}
    g=\re\lambda_1,
\end{align}
which determines the asymptotic decay rate~\citep{Kessler2012}.

Let us introduce two inner products $\braket{\hat{A},\hat{B}}$ and $\braket{\hat{A},\hat{B}}_\mathrm{ss}$ for two operators $\hat{A}$ and $\hat{B}$.
The first inner product is defined as $\braket{\hat{A},\hat{B}}=\Tr(\hat{A}^\dagger\hat{B})$.
Accordingly, we define an adjoint superoperator $\tilde{\calL}$ of $\calL$ as follows:
\begin{align}    \braket{\hat{A},\calL\hat{B}}=\braket{\tilde{\calL}\hat{A},\hat{B}}.
\end{align}
The expectation value of an Hermitian operator $\hat{A}$ at time $t$ is expressed as
\begin{align}
    \braket{\hat{A}(t)}=\Tr[\hat{A}\rho(t)]=\braket{\hat{A},e^{-\calL t}\rho}=\braket{e^{-\tilde{\calL}t}\hat{A},\rho}.
\end{align}
Here, $\hat{A}(t)\coloneqq e^{-\tilde{\calL t}}\hat{A}$ is interpreted as the time-evolved operator in the Heisenberg picture.
It is explicitly given by
\begin{align}
    -\tilde{\calL}\hat{A} =i[\hat{H},\hat{A}]+\sum_k\left(\hat{L}_k^\dagger\hat{A}\hat{L}_k-\frac{1}{2}\left\{\hat{L}_k^\dagger\hat{L}_k,\hat{A}\right\}\right).
\end{align}
Since $\tilde{\calL}$ is an adjoint of $\calL$, $\tilde{\calL}$ has the same eigenvalue spectrum as $\calL$.
We denote by $\chi_\alpha$ right eigenvectors of $\tilde{\calL}$: $\tilde{\calL}\chi_\alpha=\lambda_\alpha^*\chi_\alpha$ ($\chi_\alpha$ is also a left eigenvector of $\calL$, i.e. $\chi_\alpha^\dagger\calL=\lambda_\alpha\calL$).

The second inner product is given by
\begin{align}
\braket{\hat{A},\hat{B}}_\mathrm{ss}\coloneqq\Tr[\hat{A}^\dagger\hat{B}\rho_\mathrm{ss}],
\end{align}
which we call the steady-state inner product~\citep{Alicki1976}. 
The corresponding adjoint superoperator $\tilde{\calL}^*$ of $\tilde{\calL}$ associated with $\braket{\cdot,\cdot}_\mathrm{ss}$ is defined as
\begin{align}
\braket{\hat{A},\tilde{\calL}\hat{B}}_\mathrm{ss}=\braket{\tilde{\calL}^*\hat{A},\hat{B}}_\mathrm{ss}.
\end{align}
It should be noted that $\tilde{\calL}^*$ depends on the steady state $\rho_\mathrm{ss}$.
It is shown that $\tilde{\calL}^*$ is expressed as $\tilde{\calL}^*A=\calL(\hat{A}\rho_\mathrm{ss})\rho_\mathrm{ss}^{-1}$~\citep{Alicki1976}, where we assume that $\rho_\mathrm{ss}$ is invertible.
Again, $\tilde{\calL}^*$ has the same eigenvalue spectrum as $\calL$.
For later convenience, we define the steady-state norm $\|\cdot\|_\mathrm{ss}$ as
\begin{align}
    \|\hat{A}\|_\mathrm{ss} \coloneqq \sqrt{\braket{\hat{A},\hat{A}}_\mathrm{ss}}\geq 0.
\end{align}

\sectionprl{Main results}
Let us consider an auto-correlation function $C_A(t)=\Tr[\hat{A}(t)\hat{A}\rho_\mathrm{ss}]=\braket{\hat{A}(t),\hat{A}}_\mathrm{ss}$ in the steady state, where $\hat{A}$ is an Hermitian operator satisfying $\braket{\hat{A}}_\mathrm{ss}\coloneqq\Tr[\hat{A}\rho_\mathrm{ss}]=0$.
We investigate how quickly $C_A(t)$ decays.
It is known that the Liouvillian gap determines the asymptotic decay of $C_A(t)$:
\begin{align}
    |C_A(t)|\lesssim e^{-gt} \quad (t\to\infty).
\end{align}
However, in a transient regime, the Liouvillian gap does not necessarily give the smallest decay rate~\citep{Mori2020_resolving, Haga2021, Mori2021_metastability, Bensa2021, Lee2022}, i.e. the inequality $|C_A(t)|\leq e^{-gt}C_A(0)$ does not hold in general.
It is thus desired to give a rigorous bound on $|C_A(t)|$ at \emph{finite} times.

In this Letter, we give such a bound:
\begin{align}
    |C_A(t)|\leq e^{-g_st}C_A(0) \text{ for any }t\geq 0,
    \label{eq:main}
\end{align}
where $g_s$ is the spectral gap (i.e. the difference between the lowest and the second-lowest eigenvalues) of \emph{symmetrized Liouvillian}
\begin{align}
    \tilde{\calL}_s\coloneqq\frac{\tilde{\calL}+\tilde{\calL}^*}{2}.
\end{align}
We call $g_s$ the \emph{symmetrized Liouvillian gap}.
Because $g_s$ does not depend on $\hat{A}$, \cref{eq:main} tells us that the inverse of the symmetrized Liouvillian gap gives a general upper bound on the decay time of any auto-correlation function.

Later, we numerically show that our bound~(\ref{eq:main}) is tight in a coupled double-quantum-dot system.
It means that the symmetrized Liouvillian gap is not a mathematical artifact but a relevant quantity in the relaxation of open quantum systems.

We point out a recent work~\citep{Girotti2022} in which the symmetrized Liouvillian gap is used to derive concentration bounds for \emph{finite}-time averages of measurement outcomes in quantum Markov processes.
Such a general result is applied to derive upper bounds on the size of fluctuations of trajectory observables like time-integrated currents~\citep{Bakewell-Smith2022}, which complement lower bounds provided by thermodynamic uncertainty relations~\citep{Horowitz2020}.
In this way, the symmetrized Liouvillian gap is a key quantity to study \emph{finite}-time properties of Markov processes.

\sectionprl{Properties of $\tilde{\calL}_s$ and $g_s$}
Before proving \cref{eq:main}, we summarize basic properties of $\tilde{\calL}_s$ and $g_s$ below:
\begin{enumerate}
    \item $\tilde{\calL}_s$ has a zero eigenvalue, and $\hat{I}$ is the corresponding eigenvector.
    \item $\tilde{\calL}_s$ is positive semidefinite, i.e., all the eigenvalues are non-negative.
    \item $0\leq g_s\leq g$.
    \item $g_s=g$ when $[\tilde{\calL},\tilde{\calL}^*]=0$.
\end{enumerate}

The property (i) is easily confirmed by using $\tilde{\calL}\hat{I}=\tilde{\calL}^*\hat{I}=0$ [recall $\tilde{\calL}^*A=\calL(\hat{A}\rho_\mathrm{ss})\rho_\mathrm{ss}^{-1}$ and $\calL\rho_\mathrm{ss}=0$].

The property (ii) is proved by using the following inequality for the Liouvillain of the Lindblad form~\citep{Lindblad1976}:
\begin{align}
    \tilde{\calL}(\hat{X}^\dagger)\hat{X}+\hat{X}^\dagger\tilde{\calL}(\hat{X})\geq\tilde{\calL}(\hat{X}^\dagger\hat{X})
\end{align}
for any bounded operator $\hat{X}$.
From this inequality, we have
\begin{align}
    \braket{\hat{X},\tilde{\calL}_s\hat{X}}_\mathrm{ss}
    = \frac{1}{2}\Tr\left\{\left[\tilde{\calL}(\hat{X}^\dagger)\hat{X}+\hat{X}^\dagger\tilde{\calL}(\hat{X})\right]\rho_\mathrm{ss}\right\}
    \nonumber \\
    \geq \frac{1}{2}\Tr\left[\tilde{\calL}(\hat{X}^\dagger\hat{X})\rho_\mathrm{ss}\right]
    =\frac{1}{2}\Tr[\hat{X}^\dagger\hat{X}\calL\rho_\mathrm{ss}]=0,
\end{align}
which proves (ii).

Next, we prove (iii).
From the definition, $g_s\geq 0$ is obvious.
Because of (i) and (ii), $g_s$ is nothing but the second-lowest eigenvalue of $\tilde{\calL}_s$, which has the following variational expression:
\begin{align}
    g_s&=\inf_{\hat{X}\neq 0: \braket{\hat{X}}_\mathrm{ss}=0}\frac{\braket{\hat{X},\tilde{\calL}_s\hat{X}}_\mathrm{ss}}{\braket{\hat{X},\hat{X}}_\mathrm{ss}}
    \nonumber \\
    &=\inf_{\hat{X}\neq 0: \braket{\hat{X}}_\mathrm{ss}=0}\frac{\re\braket{\hat{X},\tilde{\calL}\hat{X}}_\mathrm{ss}}{\braket{\hat{X},\hat{X}}_\mathrm{ss}}.
    \label{eq:gs_variational}
\end{align}
The condition $\braket{\hat{X}}_\mathrm{ss}= \braket{\hat{I},\hat{X}}_\mathrm{ss}=0$ in \cref{eq:gs_variational} guarantees that $\hat{X}$ is orthogonal to the identity $\hat{I}$, which is the eigenvector of $\tilde{\calL}_s$ with zero eigenvalue.
Since $\chi_1$ satisfies $\braket{\chi_1}_\mathrm{ss}=0$ and $\tilde{\calL}\chi_1=\lambda_1^*\chi_1$, we obtain
\begin{align}
    g_s\leq\frac{\re\braket{\chi_1,\tilde{\calL}\chi_1}_\mathrm{ss}}{\braket{\chi_1,\chi_1}_\mathrm{ss}}=\re\lambda_1^*=g,
\end{align}
which proves (iii).

The following observation is key to prove the last property (iv): When $[\tilde{\calL},\tilde{\calL}^*]=0$, $\chi_\alpha$ is a simultaneous eigenvector of $\tilde{\calL}$ and $\tilde{\calL}^*$ with the eigenvalue $\lambda_\alpha^*$ and $\lambda_\alpha$, respectively.
It implies that $\chi_\alpha$ is an eigenvector of $\tilde{\calL}_s$ with the eigenvalue $\re\lambda_\alpha$.
Thus, we conclude $g_s=\min_{\alpha\neq 0}\re\lambda_\alpha=g$.

The property (iv) is of physical importance.
The condition $[\tilde{\calL},\tilde{\calL}^*]=0$ holds whenever the Liouvillian obeys the quantum detailed balance condition~\citep{Alicki1976}.
When the system is coupled to an equilibrium reservoir and its dynamics is described by the Lindblad equation with the quantum detailed balance, the standard Liouvillian gap $g$ gives a bound on the decay of any auto-correlation function as $|C_A(t)|\leq e^{-gt}C_A(0)$.
While, when the system is put in a nonequilibrium situation (e.g. the system is in contact with multiple reservoirs at different temperatures), we need $g_s$ to obtain a correct upper bound.
In this sense, $g_s$ is relevant in nonequilibrium open quantum systems.

\sectionprl{Proof of \cref{eq:main}}
We first express the auto-correlation function as $C_A(t)=\braket{\hat{A}(t),\hat{A}}_\mathrm{ss}$.
By using the Cauchy-Schwarz inequality, we obtain
\begin{align}
    |C_A(t)|\leq\|\hat{A}(t)\|_\mathrm{ss}\cdot\|\hat{A}\|_\mathrm{ss}.
    \label{eq:Schwarz}
\end{align}
Let us evaluate $f(t)\coloneqq\|\hat{A}(t)\|_\mathrm{ss}^2=\braket{\hat{A}(t),\hat{A}(t)}_\mathrm{ss}$.
By differentiating it with respect to $t$, we have
\begin{align}
    \ddv{f}{t}&=-\braket{\tilde{\calL}\hat{A}(t),\hat{A}(t)}-\braket{\hat{A}(t),\tilde{\calL}\hat{A}(t)}_\mathrm{ss}
    \nonumber \\
    &=-\braket{\hat{A}(t),(\tilde{\calL}+\tilde{\calL}^*)\hat{A}(t)}_\mathrm{ss}
    \nonumber \\
    &=-2\frac{\braket{\hat{A}(t),\tilde{\calL}_s\hat{A}(t)}_\mathrm{ss}}{\braket{\hat{A}(t),\hat{A}(t)}_\mathrm{ss}}f(t)\leq -2g_sf(t),
\end{align}
where \cref{eq:gs_variational} was used in the last inequality.
By integrating it over $t$, we obtain
\begin{align}
    f(t)\leq e^{-2g_st}f(0)=e^{-2g_st}\|\hat{A}\|_\mathrm{ss}^2.
    \label{eq:f_bound}
\end{align}
By substituting it into \cref{eq:Schwarz} and using $\|\hat{A}\|_\mathrm{ss}^2=C_A(0)$, we obtain \cref{eq:main}.

\begin{figure}
    \centering
    \includegraphics[width=0.8\linewidth]{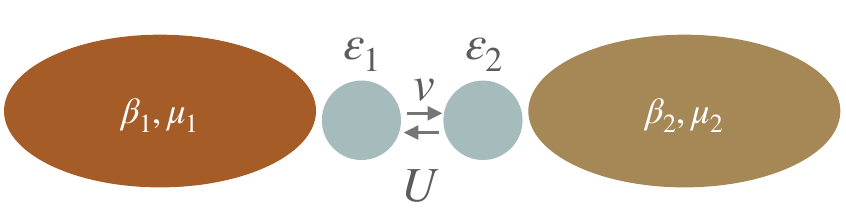}
    \caption{A schematic of the model. Two interacting quantum dots are interacting with their own reservoirs.}
    \label{fig:model}
\end{figure}

\sectionprl{Numerical results}
We demonstrate the relevance of our main results in a specific model, i.e., spinless fermions on a double quantum dot in contact with two reservoirs.
The Hamiltonian of the total system is given by $\hat{H}_T=\hat{H}_S+\hat{H}_B+\hat{H}_I$.
The Hamiltonian $\hat{H}_S$ of a double quantum dot is given by
\begin{align}
    \hat{H}_S= \sum_{i=1}^2\varepsilon_i\hat{d}_i^\dagger\hat{d}_i+v(\hat{d}_1^\dagger\hat{d}_2+\hat{d}_2^\dagger\hat{d}_1)+U\hat{d}_1^\dagger\hat{d}_1\hat{d}_2^\dagger\hat{d}_2,
\end{align}
where $\hat{d}_i$ is the annihilation operator of $i$th dot.
We denote by $E_n$ and $\ket{n}$ the energy eigenvalue and the corresponding energy eigenstate: $\hat{H}_S=\sum_nE_n\ket{n}\bra{n}$.
The Hamiltonian of the two reservoirs is given by
\begin{align}
    \hat{H}_B= \sum_k(\hat{c}_{k,1}^\dagger\hat{c}_{k,1}+\hat{c}_{k,2}^\dagger\hat{c}_{k,2}),
\end{align}
where $\hat{c}_{k,i}$ is the annihilation operator of fermions in the reservoir coupled to $i$th dot.
The interaction Hamiltonian reads
\begin{align}
    \hat{H}_I= \sum_{i=1}^2\left(\hat{d}_i^\dagger\sum_k\lambda_k\hat{c}_{k,i}+\text{h.c.}\right).
\end{align}
We assume that two reservoirs are in thermal equilibrium at the inverse temperature $\beta_i$ and the chemical potential $\mu_i$ ($i=1,2$).
See \cref{fig:model} for a schematic of the model.

When the interaction between the system and the reservoirs is sufficiently weak, the Lindblad equation for the reduced density matrix $\rho(t)$ for the system of interest is derived by applying the Born-Markov and secular approximations~\citep{Breuer_text}.
The Liouvillian is block diagonalized into sectors each of which is spanned by $\ket{n}\bra{m}$ with a fixed frequency $\omega=E_n-E_m$.
The sector of $\omega=0$ corresponds to the subspace spanned by the diagonal matrix elements $\{\ket{n}\bra{n}\}$ if we assume no energy degeneracy in $\hat{H}_S$.
Let us consider $C_A(t)$ with $\hat{A}$ being a diagonal matrix in the energy basis.
The dynamics of $\hat{A}(t)$ is then restricted to the diagonal subspace.
For this reason, we focus on the diagonal sector and define the spectral gap within this sector.

The Born-Markov-secular Lindblad equation in the diagonal sector is given by the following Pauli master equation~\citep{Breuer_text}:
\begin{align}
    \ddv{P_n}{t}=\sum_m\left[W_{nm}P_m(t)-W_{mn}P_n(t)\right],
\end{align}
where $P_n(t)=\braket{n|\rho(t)|n}$ and the transition rate matrix $W_{nm}$ is given by
\begin{align}
    W_{nm}=2\pi J(E_n-E_m)\sum_{i=1}^2f_i(E_n-E_m)\abs{\braket{n|\hat{d}_i^\dagger|m}}^2
    \nonumber \\
    +2\pi J(E_m-E_n)\sum_{i=1}^2[1-f_i(E_m-E_n)]\abs{\braket{m|\hat{d}_i|n}}^2.
\end{align}
Here, $J(\omega)=\sum_k\delta(\omega-\omega_k)|\lambda_k|^2$ is the bath spectral function and $f_i(E)=[e^{\beta_i(E-\mu_i)}+1]^{-1}$ is the Fermi distribution at $i$th reservoir.

\begin{figure}
    \centering
    \includegraphics[width=0.8\linewidth]{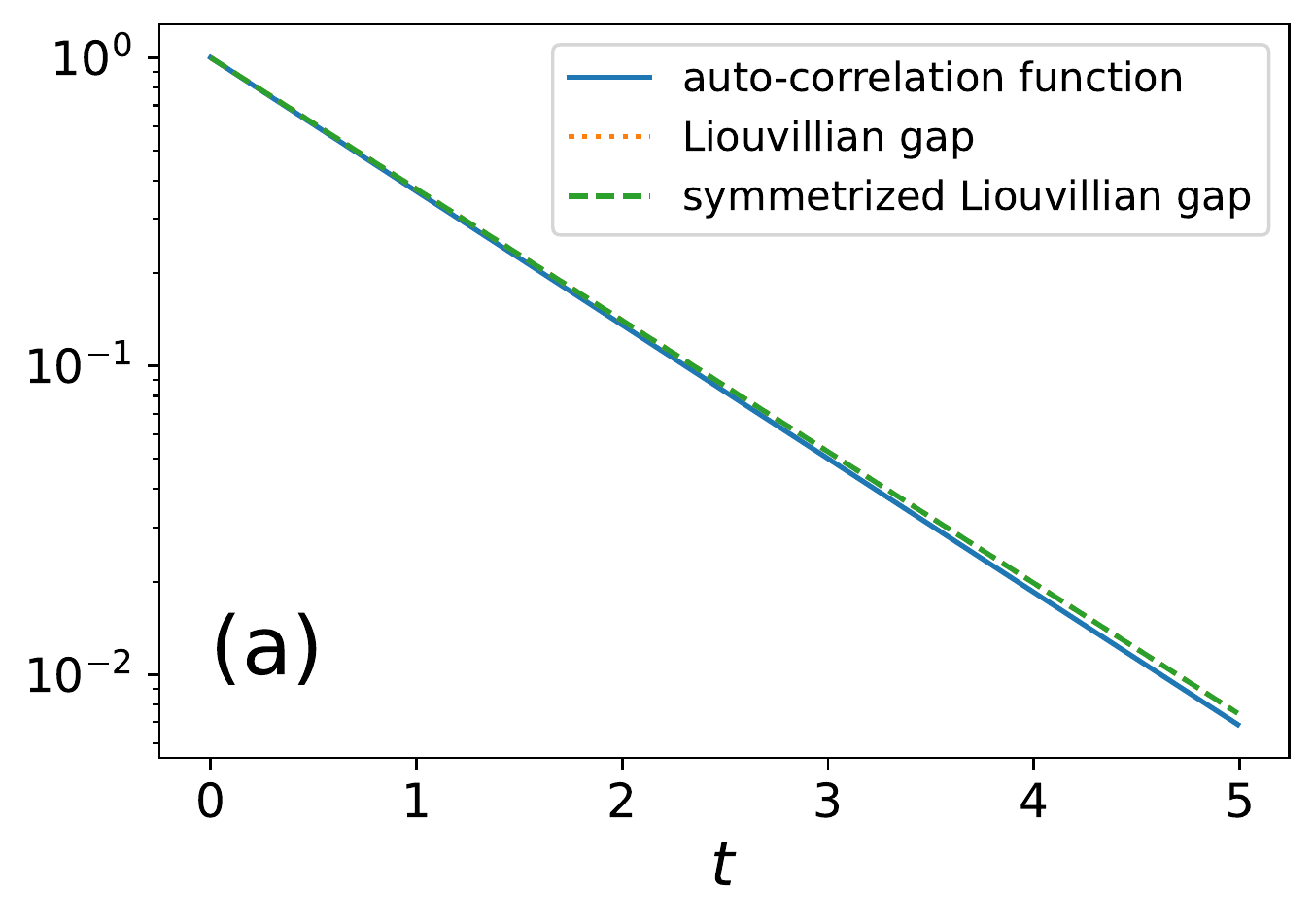}
    \includegraphics[width=0.8\linewidth]{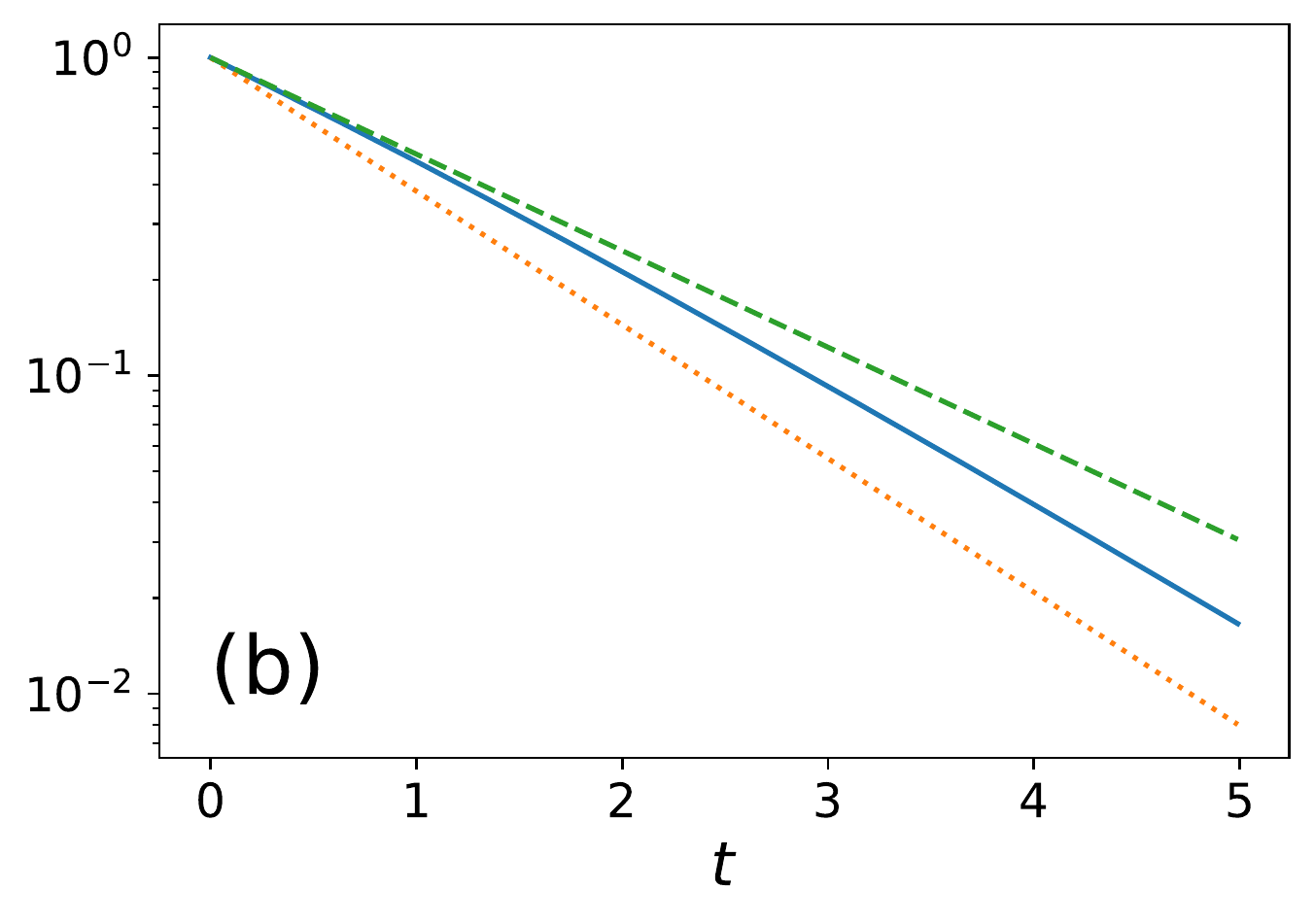}
    \caption{Auto-correlation function $C_E(t)$ for (a) equilibrium and (b) nonequiliblium cases. The solid lines are the numerical values of $C_E(t)/C_E(0)$. The orange dotted lines and the green dashed lines are $e^{-gt}$ and $e^{-g_st}$, respectively.}
    \label{fig:corr}
\end{figure}

We numerically compute the auto-correlation function of the energy,
\begin{align}
    C_E(t)=\braket{\delta\hat{H}_S(t),\delta\hat{H}_S(0)}_\mathrm{ss},
\end{align}
where $\delta\hat{H}_S=\hat{H}_S-\braket{\hat{H}_S}_\mathrm{ss}\hat{I}$.
In \cref{fig:corr}, we plot $C_E(t)/C_E(0)$, our upper bound $e^{-g_st}$, and $e^{-gt}$ for (a) an equilibrium case ($\beta_1=\beta_2$ and $\mu_1=\mu_2$) and (b) a nonequilibrium case.
In numerical calculations, we set $v=1$, $\varepsilon_1=-1.37$, $\varepsilon_2=-2.24$, $U=1.76$, and assume $J(\omega)=\gamma$ ($\gamma$ is a positive constant).
In \cref{fig:corr} (a), we set $\beta_1=\beta_2=5.5$ and $\mu_1=\mu_2=0.3$, whereas in \cref{fig:corr} (b), $\beta_1=6.94$, $\beta_2=4.06$, $\mu_1=-1.63$, and $\mu_2=2.23$.
In \cref{fig:corr} (a), we set $\beta_1=\beta_2=5.5$ and $\mu_1=\mu_2=0.3$, whereas in \cref{fig:corr} (b), $\beta_1=6.94$, $\beta_2=4.06$, $\mu_1=-1.63$, and $\mu_2=2.23$.

In an equilibrium case, $g=g_s$ and therefore the Liouvillian gap gives an upper bound on the relaxation time $\tau$ as $\tau\lesssim g^{-1}$.
While, in a nonequilibrium case, the decay rate in a transient regime is not bounded by the Liouvillian gap.
Instead, the symmetrized Liouvillian gap gives a correct bound even in this case.
Moreover, it gives a tight upper bound at short times [see \cref{fig:corr} (b)].
Thus, in general, the symmetrized Liouvillian gap is needed to evaluate the maximum relaxation time of the dissipative system.

\sectionprl{Open Floquet systems}
Our results can be extended to open Floquet systems, i.e., periodically driven dissipative quantum systems, which have been studied from long ago, but attracted renewed interests in the context of Floquet engineering in open systems~\citep{Mori2022_Floquet_review}.
An open Floquet system is described by a periodically time-dependent Liouvillian $\mathcal{L}(t)=\mathcal{L}(t+T)$, where $T$ denotes the period of the driving field.
After a sufficiently long time, the system will relax to a periodic steady state $\rho_\mathrm{ps}(t)=\rho_\mathrm{ps}(t+T)$~\citep{Ikeda2021_nonequilibrium}.

Let us fix a starting time $t_0\in[0,T)$.
The time evolution superoperator over a cycle is written as
\begin{align}
    \mathcal{U}=\mathcal{T}e^{-\int_{t_0}^{t_0+T}\mathcal{L}(t)\dd t}
    \text{ and }
    \tilde{\mathcal{U}}=\bar{\mathcal{T}}e^{-\int_{t_0}^{t_0+T}\tilde{\calL}(t)\dd t}
\end{align}
in the Schr\"odinger picture and in the Heisenberg picture, respectively, where $\mathcal{T}$ ($\bar{\mathcal{T}}$) denotes the (anti-)time-ordering operator.

We focus on the stroboscopic auto-correlation function of $\hat{A}$ that is given by
\begin{align}
    C_{A,n}\coloneqq \Tr\left[(\tilde{\mathcal{U}}^n\hat{A})\hat{A}\rho_\mathrm{ps}(t_0)\right]=\braket{\tilde{\mathcal{U}}^n\hat{A},\hat{A}}_\mathrm{ps},
\end{align}
where the inner product is defined as
\begin{align}
    \braket{\hat{A},\hat{B}}_\mathrm{ps}\coloneqq \Tr[\hat{A}^\dagger\hat{B}\rho_\mathrm{ps}(t_0)].
\end{align}
Let us denote by $\tilde{\mathcal{U}}^*$ the adjoint of $\tilde{\mathcal{U}}$: $\braket{\hat{A},\tilde{\mathcal{U}}\hat{B}}_\mathrm{ps}=\braket{\tilde{\mathcal{U}}^*\hat{A},\hat{B}}_\mathrm{ps}$.
By repeating the similar argument as in the derivation of \cref{eq:main}, we obtain
\begin{align}
    |C_{A,n}|\leq e^{-g_s(t_0)nT}C_{A,0},
\end{align}
where $g_s(t_0)$ is defined so that the second-largest eigenvalue of $\tilde{\mathcal{U}}^*\tilde{\mathcal{U}}$ is $e^{-2g_s(t_0)T}$.
In the limit of $T\to +0$, $g_s(t_0)$ is reduced to the symmetrized Liouvillian gap for the static Liouvillian $\bar{\calL}=(1/T)\int_0^T\calL(t)dt$.

\sectionprl{Summary and Outlook}
We have derived an upper bound on the decay of auto-correlation functions in the steady state, and numerically show that that bound is tight at short times.
The decay of correlations is bounded by the symmetrized Liouvillan gap, which differs from the standard Liouvillian gap only when the quantum detailed balance condition is violated.
We note that it is straithtforward to extend our results to classical Markov jump processes.

We believe that our results unveil a general property of nonequilibrium quantum dissipative dynamics.
While, we are also convinced that the symmetrized Liouvillian gap does not capture whole physics of Markovian quantum dynamics at finite times.
It is a future problem to unveil its generic properties beyond the spectral-gap analysis of the (symmetrized) Liouvillian.

\begin{acknowledgments}
This work was supported by JSPS KAKENHI Grant Numbers JP18K13466, JP19K14622, JP21H05185, and by JST, PRESTO Grant No. JPMJPR225.
\end{acknowledgments}

\bibliography{apsrevcontrol,physics}
\end{document}